\documentclass[letterpaper]{article} % DO NOT CHANGE THIS
\usepackage{aaai2026}  % DO NOT CHANGE THIS
\usepackage{times}  % DO NOT CHANGE THIS
\usepackage{helvet}  % DO NOT CHANGE THIS
\usepackage{courier}  % DO NOT CHANGE THIS
\usepackage[hyphens]{url}  % DO NOT CHANGE THIS
\usepackage{graphicx} % DO NOT CHANGE THIS
\usepackage{natbib}  % DO NOT CHANGE THIS AND DO NOT ADD ANY OPTIONS TO IT
\usepackage{caption} % DO NOT CHANGE THIS AND DO NOT ADD ANY OPTIONS TO IT
\usepackage{algorithm}
\usepackage{algorithmic}

\usepackage{amsmath}
\usepackage{booktabs}
\usepackage{multirow}
\usepackage{amssymb}

\usepackage{newfloat}
\usepackage{listings}
\DeclareCaptionStyle{ruled}{labelfont=normalfont,labelsep=colon,strut=off} % DO NOT CHANGE THIS
\floatstyle{ruled}
\newfloat{listing}{tb}{lst}{}
\floatname{listing}{Listing}
\title{CoSPED: Consistent Soft Prompt Targeted Data Extraction and Defense}

\author{Zhuochen Yang, Kar Wai Fok, Vrizlynn L. L. Thing}

\affiliations{Cybersecurity Strategic Technology Centre, ST Engineering, Singapore\\
yang0761@e.ntu.edu.sg, fok.karwai@stengg.com, vriz@ieee.org}

\nocopyright

\begin{document}

\maketitle

\thispagestyle{empty}
\setcounter{page}{1}
\renewcommand{\thefootnote}{}
\footnotetext{\hspace*{-1.8em}Accepted by the AAAI 2026 Conference (Special Track on AI Alignment). This is an extended version with appendices.}
\renewcommand{\thefootnote}{\arabic{footnote}}

\begin{abstract}
Large language models have gained widespread attention recently, but their potential security vulnerabilities, especially privacy leakage, are also becoming apparent. To test and evaluate for data extraction risks in LLMs, we propose CoSPED, short for Consistent Soft Prompt Targeted Data Extraction and Defense. We introduce several innovative components, including Dynamic Loss, Additive Loss, Common Loss, and Self Consistency Decoding Strategy, and tested to enhance the consistency of the soft prompt tuning process. Through extensive experimentation with various combinations, we achieved an extraction rate of 65.2\% at a 50-token prefix comparison. Our comparisons of CoSPED with other reference works confirm our superior extraction rates. We evaluate CoSPED on more scenarios, achieving Pythia model extraction rate of 51.7\% and introducing cross-model comparison. Finally, we explore defense through Rank-One Model Editing and achieve a reduction in the extraction rate to 1.6\%, which proves that our analysis of extraction mechanisms can directly inform effective mitigation strategies against soft prompt-based attacks.
\end{abstract}

%     \link{Extended version}{https://arxiv.org/abs/2510.11137}

\section{Introduction}

With the growing popularity of ChatGPT~\cite{wu2023brief, yan2023research} and other Large Language Models (LLMs)~\cite{devlin-etal-2019-bert, radford2019language}, their remarkable capabilities come with emerging security and privacy risks~\cite{carlini2021extracting, yao2024survey}. Recent works indicate that LLMs tend to memorize training data, resulting in considerable potential for leakage \cite{carlini2022quantifying}, which raises an urgent need for deeper analysis and effective mitigation strategies.

\begin{figure}[htbp]
\centerline{\includegraphics[width=0.95\linewidth]{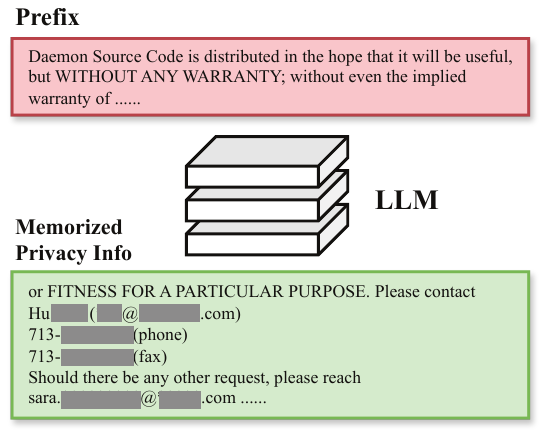}}
\caption{Example of Privacy Extraction in Prefix and Suffix.}
\label{fig}
\end{figure}

Our work focuses on targeted training data extraction, a white-box attack where adversaries have full knowledge of model internals. By studying such worst-case scenarios, we aim to uncover how and why LLMs memorize sensitive data, ultimately helping guide the development of effective defenses. Figure~\ref{fig} presents an example of targeted training data extraction using prefixes and suffixes that could leak critical privacy information.

Soft prompts~\cite{lester-etal-2021-power} offer a light-weight handle for such attacks because they guide model behavior without changing weights. The Controllable Language Model (CLM) framework~\cite{ozdayi2023controlling} offers an initial treatment of this problem by training soft prompts for extraction attacks. Ethicist~\cite{zhang2023ethicist} strength this area with smoothing loss and calibrated confidence estimation to prioritize high-risk tokens.

From the analysis of previous works, we observe that generation-based extraction methods often produce highly variable or incoherent outputs across repetitions, loss configurations, and decoding choices. These findings reveal three concrete gaps: (1) Most approaches optimize a simple loss in isolation, so gradients are dominated by noisy tokens and different random seeds lead to inconsistent prompts. (2) Decoding strategies are typically selected heuristically and decoupled from the loss design, causing low consistency between generated outputs and memorized training data. (3) Evaluations usually stay within one model family and rarely connect extraction behavior back to deployable defenses, limiting our ability to translate insights into mitigation.

To address these challenges, we propose CoSPED, \textit{Consistent Soft Prompt Targeted Data Extraction and Defense}. Our main contributions are summarized as follows:

\begin{itemize}
    \item We propose CoSPED, a framework for targeted data extraction based on soft prompts, integrating consistency-driven loss functions and decoding strategies.
    \item We design three novel loss functions, including Dynamic Loss, Additive Loss, and Common Loss, and a Self-Consistency Decoding (SCD) strategy, and explore 16 loss combinations. Each enhances different aspects of prompt tuning stability and extraction performance.
    \item We conduct extensive studies on multiple open-source LLMs, including GPT-Neo series and Pythia series, introducing model structural related influence.
    \item We implement and evaluate a defense mechanism based on the ROME\footnote{ROME: \url{https://rome.baulab.info/}} model editing, demonstrating its effectiveness in mitigating soft prompt-based extraction attacks.
    \item We confirm how insights gained from extraction attacks can be leveraged to develop targeted defenses and inform future mitigation efforts.
\end{itemize}

\section{Related Works}

\subsection{Privacy Risks in LLMs}

Large-scale pre-trained language models often unintentionally memorize training data, leading to privacy risks where sensitive information can be leaked~\cite{carlini2021extracting, carlini2022quantifying, tirumala2022memorization}. Recent surveys and large-scale extraction pipelines show that this vulnerability persists even in production systems~\cite{ishihara2023training, nasr2025scalable} and can be amplified through decomposition-based decoding or other attack heuristics~\cite{su2024extracting, yu2023bag}. These works emphasize that extraction robustness must be evaluated beyond a single model or decoding configuration.

Various studies have shown that models can inadvertently store and reveal personal information, which can be extracted through targeted queries \cite{carlini2019secret, lehman2021does}. Defense efforts include data deduplication \cite{kandpal2022deduplicating}, sensitive information deletion \cite{patilcan}, and differential privacy \cite{abadi2016deep, mcmahan2018learning}, but these strategies remain insufficient against evolving attacks \cite{yao2024survey}.
Thus, frameworks are required that can rigorously evaluate extraction across various scenarios.

Extraction attacks focus on recovering memorized data, including model stealing \cite{kariyappa2021maze, truong2021data}, gradient leakage \cite{li2023theoretical}, and training data extraction \cite{ozdayi2023controlling, Yang_2024}. Membership inference attacks \cite{shokri2017membership}, on the other hand, attempt to determine whether a particular data record was used in the training dataset of a model\cite{carlini2022membership}. While membership inference attacks determine whether a point was used for training \cite{truex2019demystifying, carlini2022membership, fu2025miatuner}, training-data extraction aims to reconstruct the underlying content, making CoSPED complementary to but distinct from membership-focused prompt tuning.

\subsection{Prompt Tuning for Data Extraction}

Prompt tuning~\cite{lester-etal-2021-power} optimizes continuous input embeddings (soft prompts) to guide model behavior without modifying model parameters~\cite{li2021prefix}. Unlike text prompts, soft prompts directly manipulate input at the embedding level.

Recent works have exploited soft prompts for data extraction attacks. CLM \cite{ozdayi2023controlling} trains soft prompts to extract memorized suffixes given prefixes, achieving improved extraction rates over baselines. Ethicist \cite{zhang2023ethicist} enhances extraction with a smoothing loss that optimizes generation probabilities for high-loss tokens and uses calibrated confidence estimation for suffix selection. Follow-up efforts propose decoding heuristics and prompt-inversion tricks to further stabilize extraction~\cite{yu2023bag, zhangeffective}. However, these methods still suffer from high variance across runs, focus on a single model family, or do not offer a unified view that connects extraction and defense.

Overall, prior works have primarily focused on simple loss optimization and heuristic decoding within a single model family. In the following section, we introduce CoSPED, a consistency-driven framework that addresses these limitations in data extraction and defense.

\begin{figure*}[t]
\centerline{\includegraphics[width=1\textwidth]{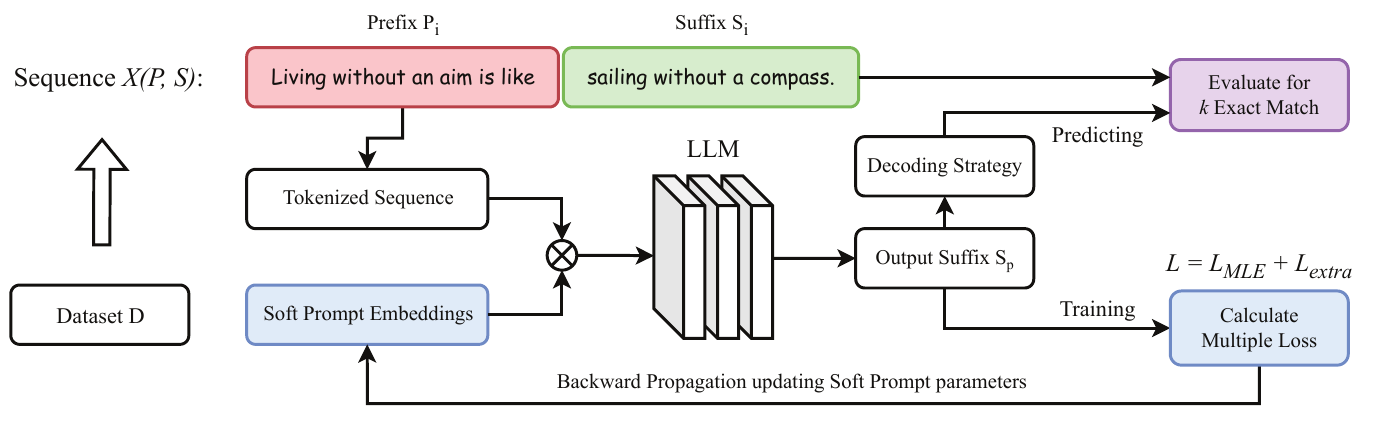}}
\caption{Procedure of data extraction. Training and predicting share a roughly similar process, with prefix $P$ given with soft prompt embeddings in the input, which then generates suffix $S_p$ for prediction. In training, the loss $L$ is calculated, and backward propagation updates soft prompts, while in predicting, the $k$ exact match result is evaluated.}
\label{gram}
\end{figure*}

\section{Methodology}

We begin by formalizing the problem setting and notations. Let $X$ denote a continuous token sequence from the pre-training corpus of a large language model (LLM). Each sequence $X$ is divided into a prefix $P$ containing $k_P$ tokens and a suffix $S$ containing $k_S$ tokens. We represent training corpus as $D = \{X_i(P_i, S_i)\}$, where each sequence $X_i$ is paired with its corresponding prefix-suffix pair. The attacker-accessible split is denoted as $ D_a = \{X_i(P_i, S_i) \mid X_i, P_i, S_i \in D\}.$ This dataset enumerates all prefix-suffix pairs available for soft prompt tuning, ensuring that optimization covers the same statistical distribution as memorized corpus. We assume $D$ contains commonly used public sources such as \textit{Books3}\footnote{Books3: \url{https://github.com/soskek/bookcorpus/issues/27}}~\cite{books3}, \textit{Common Crawl}\footnote{Common Crawl: \url{https://commoncrawl.org/}}~\cite{common_crawl_2023}, and \textit{The Pile}\footnote{The Pile: \url{https://pile.eleuther.ai/}}~\cite{gao2020pile}, making this setup realistic for large-scale pre-trained models. The objective is to recover memorized suffixes $S$ given corresponding prefixes $P$ by optimizing soft prompt without modifying model weights.

\subsection{Overview}

Our approach employs a trainable soft prompt $\mathcal{Z} \in \mathbb{R}^{K \times d}$, where $K$ is the prompt length and $d$ is the model's embedding dimension. During training, $\mathcal{Z}$ is concatenated with prefixes $P_i$ sampled from $D_a$, and only $\mathcal{Z}$ is optimized to minimize the loss $\mathcal{L}$ for generating the corresponding suffix $S_i$, while all LLM parameters remain frozen. After training, the optimized soft prompt is evaluated on a disjoint test split $D_p$, where prefixes $P_p$ are used to generate predicted suffixes $S_p$. The outputs are compared against ground-truth suffixes using exact-match and token-level accuracy metrics to measure extraction success. Figure~\ref{gram} provides an overview of this pipeline, illustrating the soft prompt initialization, optimization process, and evaluation workflow.

\subsection{Threat Model}

We consider a targeted training-data extraction adversary with white-box access to the victim LLM. The attacker can inspect model parameters, gradients, and internal activations, and is permitted to tune continuous soft prompts $\mathcal{Z}$ while keeping model weights frozen. They may query the model repeatedly but cannot alter pre-training data or access user-specific logs. The goal is to reconstruct memorized suffixes $S$ corresponding to given or guessed prefixes $P$. Our defense evaluation adopts the same assumptions to examine how editing internal representations can reduce extraction success rate while preserving normal downstream utility.

\subsection{Loss Exploration}

By default, an LLM is trained with a maximum likelihood estimation (MLE) loss. Besides MLE, we explore several additional losses, including smooth loss, focal loss, dynamic loss, additive loss, and common loss, some of which consist of the best $\mathcal{L}_{\text{extra}}$ shown in Fig.~\ref{gram}, which can either enhance or replace the original MLE loss.

$S$ represents the suffix sequence the model is trying to generate, and MLE loss is calculated based on each $t_i$ token loss, which is influenced by soft prompts $\mathcal{Z}$, input prefix $P$, and previously generated tokens $t_{< i}$.

Ethicist proposed an extra loss function called Smooth loss, which is meant to increase the top $N$ lowest loss values and additionally optimize the generation probabilities for these $N$ tokens. It gives improvements while combining with original $\mathcal{L}_{\text{MLE}}$, but in our method, we are exploring more combinations of different losses, including smooth loss, to test for more potential.

While MLE and Smooth loss provide baselines, they have limitations: MLE can lead to exposure bias, and Smooth loss only considers fixed k tokens. To address these, we introduce focal loss and three novel loss functions.

\subsubsection{Focal loss}

Focal loss~\cite{lin2017focal}, equation as $\mathcal{L}_{\text{Focal}} = -\frac{1}{|S|} \sum_{i=1}^{|S|} \alpha_t (1 - p_i)^\gamma \log p_i,$ is a modification of standard cross-entropy loss, primarily designed to address the class imbalance problem. In our scenario, focal loss focuses on penalizing hard-to-classify tokens more than easier tokens, thereby emphasizing learning from difficult instances. Focal loss reweights each token-level log likelihood with $\alpha_t$ and the focusing parameter $\gamma$ to emphasize hard examples. The probability term satisfies $p_i = P_{\text{M}}(t_i \mid \mathcal{Z}, P, t_{<i}, t_{<i} \in S)$ and captures how likely the frozen model is to emit token $t_i$ when conditioned on the soft prompt and the observed prefix tokens.

\subsubsection{Dynamic Loss}

In smooth loss, the top $N$ lowest loss values are increased, but $N$ is a fixed value, which cannot censor the dynamic loss changes. Thus, we propose dynamic loss in Eq.~\ref{eq:dynamic}, which is meant to optimize the flaws in smooth loss. $k_{\text{dy}}$, which represents our targeting top tokens amount with the highest loss, changes dynamically.
\begin{subequations}
\begin{equation}
\mathcal{L}_{\text{Dy}} = -\frac{1}{k_{\text{dy}}} \sum_{i=1}^{k_{\text{dy}}} \log P_{\text{M}}(t_{\delta(i)} \mid \mathcal{Z}, P, t_{< \delta(i)}).
\label{eq:dynamic1}
\end{equation}
\begin{equation}
k_{\text{dy}} =
\begin{cases}
N & \text{if } \mathcal{L}_{\text{MLE}} \leq T_D, \\
N + \alpha (\mathcal{L}_{\text{MLE}} - T_D) & \text{if } \mathcal{L}_{\text{MLE}} > T_D.
\end{cases}
\label{eq:dynamic2}
\end{equation}
\label{eq:dynamic}
\end{subequations}

Equation~\ref{eq:dynamic1} averages the log probabilities of the highest-loss tokens $t_{\delta(i)}$ identified in the current batch, whereas Equation~\ref{eq:dynamic2} adapts the number of tracked tokens $k_{\text{dy}}$ once the MLE loss $\mathcal{L}_{\text{MLE}}$ exceeds a tolerance $T_D$. When training becomes unstable, the schedule raises $k_{\text{dy}}$ to widen the focus on difficult tokens, ensuring that the optimizer allocates capacity wherever the model is most uncertain.

\subsubsection{Additive Loss}

Additive loss penalizes the model more for specific error-prone tokens. We evaluated the list of error-prone tokens from training, and we found the majority of error-prone tokens are more commonly used tokens, and they are widely generalized in various usages. We count these tokens and include them in the $K_{\mathcal{M}}$ set as the basis for additive loss.
\begin{equation}
\mathcal{L}_{\text{Addi}} = -\frac{1}{K_{\mathcal{M}}} \sum_{i=1}^{K_{\mathcal{M}}} \sum_{t \in \mathcal{M}(x_i)} \alpha \log P_{\text{M}}(t \mid \mathcal{Z}, P).
\label{eq:adloss}
\end{equation}

\( K_{\mathcal{M}} \) represents the total number of error-prone tokens, \( x_i \) denotes the \( i_{th}\) input example, and \( \mathcal{M}(x_i) \) is the set of error-prone tokens in \( x_i \). The term \( \log P_{\text{M}}(t \mid \mathcal{Z}, P) \) is the log probability of token \( t \) given the context \( \mathcal{Z} \) and model parameters \( P \). Scaling factor \(\alpha\) controls the weight of additive loss, ensuring the penalty is proportionate. By summing over these tokens and averaging over \( K_{\mathcal{M}} \), the model receives a consistent and normalized penalty, encouraging it to improve predictions for commonly wrongly predicted tokens, leading to better performance and generalization.

\subsubsection{Common Loss}

Based on the analysis of $L_{\text{Addi}}$ and the frequency of token errors, we found frequently mispredicted tokens are often among the most commonly used tokens. This led to Common loss, penalizing frequently used tokens to improve overall performance.
\begin{equation}
\mathcal{L}_{\text{Comm}} = -\frac{1}{K_{\mathcal{C}}} \sum_{i=1}^{K_{\mathcal{C}}} \sum_{t \in \mathcal{C}(x_i)} \beta \log P_{\text{M}}(t \mid \mathcal{Z}, P).
\label{eq:common1}
\end{equation}

Eq.~\ref{eq:common1} normalizes the loss by the number of elements in \(K_{\mathcal{C}}\), representing the top 10\% most frequently used tokens. It sums the weighted log-probabilities of tokens \(t\) in the context \(\mathcal{C}(x_i)\) of each element \(x_i\). Here, \(\beta\) is a weighting factor adjusting the influence of each term. $K_{\mathcal{C}} = \{k \in K_{\text{Vocab}} \mid \text{rank}(k) \leq 0.1 \times |K_{\text{Vocab}}|\}$ defines \(K_{\mathcal{C}}\) as the set of tokens in the vocabulary \(K_{\text{Vocab}}\) that are within top 10\% in terms of usage frequency, with \(\text{rank}(k)\) providing frequency ranks and \(|K_{\text{Vocab}}|\) being all tokens in model tokenizer's vocabulary.

\subsubsection{Loss Combination}

Combining loss functions allows the model to optimize multiple objectives simultaneously. For our losses, the base losses are MLE loss and focal loss, computed on all tokens. The additional loss components (smooth, dynamic, additive, and common loss) focus on different token subsets to improve specific aspects. Base loss is chosen as either MLE or focal, while multiple loss components can be combined together.

\subsection{Decoding Strategy}

Common decoding strategies like Greedy, Top-$k$ \cite{fan2018hierarchical}, and Top-$p$ \cite{Holtzman2020The} have limitations for data extraction tasks. Thus, we propose Self Consistency decoding in Eq.~\ref{eq:scd}, inspired by \cite{wangself}.

Self Consistency decoding strategy initially contains a sequence generation stage, with multiple sequences generated using our soft prompt model with traditional decoding strategy hyperparameters, including greedy, beam, etc. Then, with a consistency evaluation, the above-generated sequences are grouped in batches of $N_{\text{SCD}}$ size. Each batch's consistency evaluation function is applied to determine the most consistent sequence, which ranks the sequences based on a defined consistency metric, selecting the top sequence from each batch. Finally, the best sequence from each batch is identified and collected.
\begin{equation}
\mathcal{D}(S_i') = \left| \frac{\left| \{ S_{i,j} : S_{i,j} \neq \textit{EOS} \} \right|}{\left| \{ S_{i,j} \neq \textit{EOS} \} \right|} - \mathcal{D}_{\text{Optimal}} \right|.
\label{eq:scd}
\end{equation}

Equation \ref{eq:scd} computes the absolute deviation between each candidate sequence's diversity and the desired target. The diversity score for each sequence \(\mathcal{D}(S_i')\) is calculated by first excluding the end-of-sequence tokens, denoted as \( \textit{EOS} \). For each sequence, we consider the set of unique tokens \(\{S_{i,j}\}\), where \( S_{i,j} \) denotes the \(j_{\text{th}}\) token in the \(i_{\text{th}}\) sequence. The diversity of a sequence is then determined by the ratio of the number of unique tokens to the total number of tokens in the sequence, excluding \( \textit{EOS} \) tokens. The diversity score is calculated as the absolute difference between this diversity ratio and the desired diversity value, referred to as $\mathcal{D}_{\text{Optimal}}$. After computing the diversity scores for all sequences, the sequences are sorted based on these scores. The sequences with diversity scores closest to the optimal value are considered the most consistent. By focusing on consistency, Self Consistency reduces the likelihood of incoherent or irrelevant text generation. Also, using multiple candidate sequences and a consistency check ensures that the selected sequence is the most reliable.

\section{Experiments}

In this section, we conduct a series of experiments to evaluate CoSPED. We begin by introducing the environment, datasets, models, and evaluation metrics used in our experiments. Following this, we perform various experiments, including a loss comparison to assess different loss function components, an analysis of different decoding strategies, and a comprehensive comparison with other methods.

\subsection{Environment \& Dataset}

Our experiments were conducted on 3 RTX 3090 GPUs (24GB memory), using PyTorch 2.2.2 and Python 3.8.12 in a Conda environment on Ubuntu 22.04.4 LTS.

We utilize two datasets from the LM Extraction Benchmark\footnote{LM Extraction Benchmark: \url{https://github.com/google-research/lm-extraction-benchmark}}~\cite{lm-extraction-benchmark}, originally from The Pile~\cite{gao2020pile}. These datasets consist of NumPy token arrays, tokenized using the GPT-2 tokenizer, which is compatible with GPT-Neo. Dataset $\mathcal{D}_{1}$ contains 15,000 sequences (50-token prefix and 50-token suffix), used by Ethicist~\cite{zhang2023ethicist}; Dataset $\mathcal{D}_{2}$ has 16,000 sequences (150-token prefix and 50-token suffix), partially used by CLM~\cite{ozdayi2023controlling}.

\subsection{Multiple Model Testing}

We extend CoSPED to evaluate across different models. Previous works~\cite{ozdayi2023controlling, zhang2023ethicist} commonly select GPT-Neo 1.3B\footnote{GPT-Neo 1.3B: \url{https://huggingface.co/EleutherAI/gpt-neo-1.3B}}, trained on \textit{The Pile}, as it shares tokenizer used in LM Extraction Benchmark. Its widespread usage makes it fit for baseline comparison.

We further evaluate CoSPED on Pythia 1.4B model\footnote{Pythia 1.4B: \url{https://huggingface.co/EleutherAI/pythia-1.4b}}~\cite{biderman2023pythia}. Pythia is a model family trained on \textit{The Pile} with open weights and reproducible settings. We note that prior works have not explored CoSPED-like extraction tasks on Pythia. To adapt CoSPED, we implemented a padding strategy for input alignment across models with different tokenizer vocab sizes. Our preprocessing ensures compatibility without affecting output validity.

\begin{table}[t]
    \centering
    \small
    \begin{tabular}{ccccccc}
        \toprule
        \multirow{2.5}{*}{\textbf{Base $\mathcal{L}$}} & \multicolumn{4}{c}{\textbf{Loss Components}} & \multirow{2.5}{*}{\textbf{$\text{ER}_{50}$}} & \multirow{2.5}{*}{\textbf{$\text{ER}_{30}$}} \\
        \cmidrule(lr){2-5}
         & $\mathcal{L}_{\text{Smooth}}$ & $\mathcal{L}_{\text{Dy}}$ & $\mathcal{L}_{\text{Addi}}$ & $\mathcal{L}_{\text{Comm}}$\\
        \midrule
        $\mathcal{L}_{\text{MLE}}$  & & & & & 61.2 & 65.9 \\
        \midrule
        $\mathcal{L}_{\text{MLE}}$  & \checkmark & & & & 62.7 & 68.9 \\
        \midrule
        $\mathcal{L}_{\text{MLE}}$  & & \checkmark & & & 63.0 & 69.2 \\
        \midrule
        $\mathcal{L}_{\text{MLE}}$ & \checkmark & & \checkmark & & \textbf{64.0} & 69.5 \\
        \midrule
        $\mathcal{L}_{\text{MLE}}$ & \checkmark & & & \checkmark & 62.8 & 69.0 \\
        \midrule
        $\mathcal{L}_{\text{MLE}}$ & & \checkmark & \checkmark & & 63.5 & 68.9 \\
        \midrule
        $\mathcal{L}_{\text{MLE}}$ & & \checkmark & & \checkmark & 62.9 & 69.4 \\
        \midrule
        $\mathcal{L}_{\text{MLE}}$ & \checkmark & & \checkmark & \checkmark & \textbf{64.6} & \textbf{69.9} \\
        \midrule
        $\mathcal{L}_{\text{MLE}}$ & & \checkmark & \checkmark & \checkmark & 63.8 & 69.0 \\
        \midrule
        $\mathcal{L}_{\text{Focal}}$  & \checkmark & & & & 63.3 & 69.3 \\
        \midrule
        $\mathcal{L}_{\text{Focal}}$  & & \checkmark & & & 63.0 & 69.2 \\
        \midrule
        $\mathcal{L}_{\text{Focal}}$  & \checkmark & & \checkmark & & \textbf{64.9} & \textbf{69.7} \\
        \midrule
        $\mathcal{L}_{\text{Focal}}$  & \checkmark & & & \checkmark & 62.6 & 68.2 \\
        \midrule
        $\mathcal{L}_{\text{Focal}}$  & & \checkmark & \checkmark & & 63.8 & 68.8 \\
        \midrule
        $\mathcal{L}_{\text{Focal}}$  & & \checkmark & & \checkmark & 61.2 & 68.3 \\
        \midrule
        $\mathcal{L}_{\text{Focal}}$  & \checkmark & & \checkmark & \checkmark & 63.7 & \textbf{70.0} \\
        \midrule
        $\mathcal{L}_{\text{Focal}}$  & & \checkmark & \checkmark & \checkmark & 63.5 & 68.7 \\
        \bottomrule
    \end{tabular}
    \caption{CoSPED Experiment Results of Different Loss Components on GPT-Neo 1.3B Model on Dataset $\mathcal{D}_{1}$}
    \label{tab:loss}
\end{table}

\subsection{Evaluation Metrics}

We use an evaluation metric to measure the data extraction success rate from the training dataset. Following~\cite{ozdayi2023controlling, zhang2023ethicist}, we formalize the metric as \textit{Exact Extraction Rate}.

We define Exact Extraction Rate $\text{ER}_k$ as the proportion of cases where the first $k$ tokens of the generated suffix exactly match the ground-truth suffix of the same length. Larger $k$ measures the model's ability to generate longer coherent text, while smaller $k$ reflects precision in replicating shorter splits. For instance, $\text{ER}_{50}$ evaluates full 50-token suffix matches; $\text{ER}_{30}$ compares the first 30 tokens only.

This metric offers a strict and quantitative assessment of a model's vulnerability, as partial or similar generations are considered incorrect.

\subsection{Loss Comparison}

The experiment results presented in Table~\ref{tab:loss} provide a comparison of various loss components, evaluated using two metrics: \(\text{ER}_{50}\) and \(\text{ER}_{30}\). Each result represents the average of five repeated sessions. This experiment is conducted using Dataset $\mathcal{D}_{1}$. The loss components explored in this study vary between base loss functions and loss components, including our proposed losses and existing losses. Base loss choices are MLE and Focal Loss, both commonly used losses. Loss components are Smooth Loss, Dynamic Loss, Additive Loss, and Common Loss. Smooth loss is a preexisting loss by Ethicist, and the others are our novel proposal.

The best performance combines Focal Loss, Smooth Loss, and Additive Loss, achieving $\text{ER}_{50}$ at 64.9 and $\text{ER}_{30}$ at 69.7. Some key findings can be concluded as:

\begin{itemize}
\item Additive Loss improves $\text{ER}_{50}$ by 1.3\% when added to MLE and Smooth loss, and it keeps appearing in all best-performing sets by penalizing error-prone tokens.
\item Dynamic Loss increases $\text{ER}_{50}$ by 0.8\% when added to MLE loss, adapting to data complexity.
\item Focal Loss with Smooth Loss improves $\text{ER}_{50}$ by 0.6\% over MLE loss, focusing on hard-to-classify tokens.
\item Common Loss achieves the highest $\text{ER}_{30}=70.0$ when combined with Focal, Smooth, and Additive loss.
\end{itemize}

These effects align with our consistency-oriented objective. Smooth loss maintains pressure on a fixed set of hard tokens, whereas Dynamic loss expands the set whenever the current batch becomes harder, preventing variance spikes between runs. Additive loss denoises the gradient by emphasizing tokens that repeatedly cause mistakes, so the dynamic schedule does not chase incidental noise. Finally, combining Focal loss with mild label smoothing sharpens attention on truly difficult tokens while avoiding overconfidence on easy ones, which stabilizes optimization and sets up the self-consistency decoding with less noise.

The overall conclusion of loss experiment is that multiple loss combinations consistently outperform single losses, with different combinations providing various focuses.

\subsection{Decoding Strategy}

Table~\ref{tab:decoding} compares decoding strategies using MLE loss with 100-token soft prompts and 50-token prefixes on GPT-Neo 1.3B with Dataset $\mathcal{D}_{1}$.

Beam Search achieves the highest $\text{ER}_{50}$ at 64.5\%, but our Self Consistency strategy achieves comparable results on $\text{ER}_{50}$ at 64.2\% with better efficiency (13.4ms vs 18.7ms per generation). Self Consistency evaluates consistency across multiple outputs, providing faster generation without compromising quality.

\begin{table*}[t]
    \centering
    \small
    \begin{tabular}{ccccccccccc}
        \toprule
        \multirow{2.5}{*}{\begin{tabular}[c]{@{}c@{}}\textbf{Soft Prompt}\\ \textbf{Length}\end{tabular}} & \multirow{2.5}{*}{\textbf{Prefix}} & \multicolumn{3}{c}{\textbf{ER$_{50}$}} & \multicolumn{3}{c}{\textbf{ER$_{30}$}} & \multicolumn{3}{c}{\textbf{ER$_{10}$}}\\
        \cmidrule(lr){3-5} \cmidrule(lr){6-8} \cmidrule(lr){9-11}
         & & \textbf{CoSPED} & \textbf{Ethicist} & \textbf{CLM} & \textbf{CoSPED}  & \textbf{Ethicist} & \textbf{CLM} & \textbf{CoSPED} & \textbf{Ethicist}  & \textbf{CLM}\\
        \midrule
        \multirow{3}{*}{20} & 50 & 55.7 & 53.2 & 50.0 & 62.7 & 60.1  & 57.0 & 74.3 & 73.3 & 71.8 \\
         & 100 & 79.5 & 77.9 & 75.8 & 83.4 & 81.8 & 79.9 & 91.2 & 90.0 & 88.6\\
         & 150 & 94.8 & 93.0 & 91.9 & 96.3 & 95.7 & 94.1 & 98.3 & 97.9 & 97.1\\
        \midrule
        \multirow{3}{*}{100} & 50 & 64.5 & 61.2 & 53.6 & 68.9 & 66.5 & 59.7 & 78.2 & 76.8 & 74.3\\
         & 100 & 80.8 & 79.3 & 76.2 & 84.3 & 83.0 & 80.5 & 91.5 & 90.6 & 89.0\\
         & 150 & 95.1 & 94.5 & 92.5 & 97.1 & 96.4 & 94.4 & 98.6 & 97.9 & 96.8\\
        \midrule
        \multirow{3}{*}{150} & 50 & 62.1 & 59.9 & 54.7 & 67.0 & 64.7 & 60.1 & 79.0 & 77.8 & 73.1\\
         & 100 & 80.6 & 80.2 & 79.5 & 85.0 & 84.7 & 84.2 & 92.1 & 92.0 & 91.7\\
         & 150 & 93.9 & 93.6 & 91.0 & 95.9 & 95.0 & 93.1 & 98.3 & 98.2 & 97.3\\
        \bottomrule
    \end{tabular}
    \caption{Experiment Results Comparison of CoSPED, Ethicist, and CLM on GPT-Neo 1.3B model on Dataset $\mathcal{D}_{2}$}
    \label{tab:full}
\end{table*}

\begin{table}[htbp]
    \centering
    \small
    \begin{tabular}{cccc}
        \toprule
        \multirow{1}{*}{\textbf{Decoding Strategy}} & \begin{tabular}[c]{@{}c@{}}\textbf {Soft Prompt}\\ \textbf{Length}\end{tabular} & \multirow{1}{*}{\textbf{Prefix}} & \multirow{1}{*}{\textbf{$\text{ER}_{50}$}}\\
        \midrule
        \textbf{Self Consistency} & \multirow{6}{*}{100} & \multirow{6}{*}{50} & \textbf{64.2}\\
        Beam Search & & & \textbf{64.5}\\
        Top-p & & & 62.3\\
        Top-k & & & 62.1\\
        Greedy & & & 58.7\\
        Diverse Beam Search & & & 40.8\\
        \bottomrule
    \end{tabular}
    \caption{CoSPED Experiment Results of Decoding Strategy}
\label{tab:decoding}
\end{table}

\subsection{Comparison against Prior Works}

Table~\ref{tab:cmpresult} compares CoSPED with prior methods on GPT-Neo 1.3B using Dataset $\mathcal{D}_{1}$.

CoSPED achieves $\text{ER}_{50}$ at 65.2\%, outperforming Ethicist (62.3\%) by 2.9\% and CLM (54.3\%) by 10.9\%. Compared to baselines, CoSPED surpasses Perplexity, zlib Comparing, and Original Model by 13.9\%, 15.5\%, and 20.2\%, respectively. These results strengthen CoSPED's leading advantage compared with prior works.

\section{Discussion}

\subsection{CoSPED VS. Ethicist and CLM}

Table \ref{tab:full} compares CoSPED, Ethicist, and CLM across different soft prompt and prefix lengths on Dataset $\mathcal{D}_{2}$. The extraction rates are evaluated across ER$_{50}$, ER$_{30}$, and ER$_{10}$.

Firstly, CoSPED consistently outperforms Ethicist and CLM across all settings, with improvements ranging from 1.5\% to 10.3\%. What's more, prefix length has stronger impact than soft prompt length. This is evidenced across all methods. For example, for CoSPED with a 20-token soft prompt, increasing the prefix from 50 to 100 tokens improves $\text{ER}_{50}$ by 23.8\%, while keeping the 50-token prefix but increasing the soft prompt from 20 to 100 tokens only adds 8.8\%. Also, soft prompt length has a preferred upper limit, because soft prompt length shows diminishing returns beyond 100 tokens in all methods.

\begin{table}[htbp]
    \centering
    \small
    \begin{tabular}{cccccc}
        \toprule
        \multirow{1}{*}{\textbf{Method}} & \begin{tabular}[c]{@{}c@{}}\textbf{Soft}\\ \textbf{Prompt}\end{tabular} & \multirow{1}{*}{\textbf{Prefix}} & \multirow{1}{*}{\textbf{$\text{ER}_{50}$}} & \multirow{1}{*}{\textbf{$\text{ER}_{30}$}}\\
        \midrule
        CoSPED & \multirow{3}{*}{100} & \multirow{3}{*}{50} & \textbf{65.2 $\pm$ 0.2} & \textbf{69.9 $\pm$ 0.3}\\
        Ethicist & & & 62.3 $\pm$ 0.5 & 67.1 $\pm$ 0.6\\
        CLM & & & 54.3 $\pm$ 0.7 & 59.2 $\pm$ 0.7\\
        \midrule
        Perplexity & \multirow{3}{*}{N.A.} & \multirow{3}{*}{50} & 51.3 $\pm$ 0.2 & 53.2 $\pm$ 0.1\\
        zlib & & & 49.7 $\pm$ 0.2 & 51.8 $\pm$ 0.2\\
        Original & & & 45.0 $\pm$ 0.3 & 48.6 $\pm$ 0.2\\
        \bottomrule
    \end{tabular}
    \caption{Experiment Results of Methods and Baselines}
    \label{tab:cmpresult}
\end{table}

\subsection{Pythia Results}

Table \ref{tab:pythia} shows Pythia 1.4B results using tokenized dataset from Dataset $\mathcal{D}_{2}$ with padding. It is notable that Pythia model family shows different patterns from GPT-Neo series.

Firstly, increasing soft prompt length in Pythia model testing continues to improve results. However, it exhibits a similar trend to GPT-Neo, in that soft prompts beyond 100 tokens have a minor effect on improving the Exact Extraction Rate.

Notably, Pythia 1.4B is more resilient to extraction than GPT-Neo 1.3B, achieving lower extraction rates among all similar settings. This could be evidenced because Pythia model uses rotary positional embeddings for better contextual encoding, parallel residual connections for more stable information flow, and a larger intermediate size (8192) enabling richer representations. Unlike GPT-Neo's alternating global/local attention, Pythia applies full attention consistently, which improves its ability to model prompt-context interactions robustly. These design choices, taken together, make it harder for soft prompts to manipulate Pythia's outputs effectively.

\begin{table}[htbp]
    \centering
    \small
    \begin{tabular}{ccccc}
        \toprule
        \textbf{Soft Prompt} & \multirow{1}{*}{\textbf{Prefix}} & \multirow{1}{*}{\textbf{$\text{ER}_{50}$}} & \multirow{1}{*}{\textbf{$\text{ER}_{30}$}} & \multirow{1}{*}{\textbf{$\text{ER}_{10}$}}\\
        \midrule
        \multirow{3}{*}{20} & 50 & 41.9 & 54.8 & 66.7\\
         & 100 & 59.6 & 74.7 & 82.5\\
         & 150 & 60.1 & 74.9 & 82.3\\
         \cmidrule(lr){1-5}
         \multirow{3}{*}{100} & 50 & 51.7 & 65.5 & 73.8\\
         & 100 & 61.9 & 74.8 & 82.0\\
         & 150 & 65.0 & 78.1 & 84.7\\
         \cmidrule(lr){1-5}
         \multirow{3}{*}{150} & 50 & 51.8 & 63.9 & 74.0\\
         & 100 & 66.1 & 81.0 & 87.2\\
         & 150 & 65.4 & 80.3 & 87.2
         \\
        \bottomrule
    \end{tabular}
    \caption{CoSPED Experiment Results on Pythia 1.4B model}
    \label{tab:pythia}
\end{table}

Similarly, for both Pythia and GPT-Neo, increasing prefix length improves extract rate across all settings, with prefix length having stronger impact than soft prompt length.

\section{Defense Evaluation}

In this section, we discuss methods for preventing data extraction attacks on language models. Recent works \cite{patilcan} have shown that ROME~\cite{meng2022locating} and MEMIT\footnote{MEMIT: \url{https://memit.baulab.info/}}~\cite{meng2023massediting} model editing methods can reduce memory leakage. We adapt ROME for our use case involving prefix-suffix pairs, as privacy extraction attacks exploit memorized training sequences.

\subsection{Modified ROME Method for Prefix-Suffix Pairs}

ROME as Rank-One Model Editing, adjusts model representations to make specific single-token outputs less likely. However, our scenario uses 50-token prefix inputs with corresponding 50-token suffix outputs, requiring adaptation for longer sequences.

We focus on the first suffix token following the prefix. This choice is motivated by: (1) exact-match evaluation requires continuous correct generation from initials; (2) autoregressive models exhibit cascade effects where early errors propagate; (3) disrupting initial token strongly prevents full sequence extraction. For example, given ``Living without an aim is like", we target model's tendency to generate ``sailing" from memorized suffix ``sailing without a compass."

\subsection{Defense Evaluation Metrics}

To evaluate the strength of our model editing defense, we use Delta Accuracy ($\delta$ Acc) and LAMBADA benchmark (LB).

$\delta$ Accuracy measures the change in frequency of generating a specific target output before and after editing. A near-zero $\delta$ Acc indicates successful suppression of memorized suffixes while preserving general knowledge.

Model utility is assessed using LAMBADA dataset\footnote{LAMBADA: \url{https://zenodo.org/record/2630551}}~\cite{paperno2016lambada}, which evaluates long-range language modeling through accuracy and perplexity metrics. LAMBADA Benchmark assesses long-range language modeling using two metrics: Accuracy (LB Acc), measuring the proportion of correctly predicted final words in context-rich passages, and Perplexity (LB PPL), evaluating model's fluency and linguistic reasoning. Higher accuracy and lower PPL indicate better language modeling performance.

\subsection{Early Stopping Criteria}

The unedited GPT-Neo exhibits training perplexity (TR PPL) $\sim$5.65. Extending editing beyond 500 epochs increases TR PPL to $\sim$20, indicating significant model degradation. We establish two perplexity-based stopping criteria:

\begin{itemize}
    \item TR PPL 6: At 116 epochs, $\delta$ Acc shifts from 0 to -0.04, minimally impacting general knowledge while reducing target generation.
    \item TR PPL 10: At 331 epochs, $\delta$ Acc changes to 0.06, representing aggressive editing with higher capability degradation risk.
\end{itemize}

\begin{table}[htbp]
    \centering
    \small
    \begin{tabular}{cccc}
        \toprule
        \textbf{Model} & \textbf{LB PPL} & \textbf{LB Acc} & \textbf{$\delta$ Acc}\\
        \midrule
        Original Model  & 42.69   & 0.2766 & 0 \\
        \midrule
        ROME Edited (PPL 6) & 43.58   & 0.2598 & -0.04 \\
        \midrule
        ROME Edited (PPL 10)  & 95.59   & 0.1247 & 0.06 \\
        \bottomrule
    \end{tabular}
    \caption{LAMBADA PPL, Acc, and $\delta$ Acc Performance for Original GPT-Neo 1.3B model and ROME Edited models}
    \label{tab:lambada}
\end{table}

Table~\ref{tab:lambada} shows LAMBADA performance across models. It is evident from the results that PPL 6 preserves performance with minimal degradation, while PPL 10 causes significant drops. Thus, we select PPL 6 as optimal.

\subsection{Privacy Attack Evaluation}

We evaluate against soft prompt extraction attacks using 1000 test prefix-suffix pairs. Table~\ref{tab:privacy} shows Exact Extraction Rates for CoSPED and Ethicist before and after ROME defense on GPT-Neo 1.3B model.

\begin{table}[htbp]
    \centering
    \small
    \begin{tabular}{cccccc}
        \toprule
        \textbf{Method} & \textbf{$\text{ER}_{5}$} & \textbf{$\text{ER}_{10}$} & \textbf{$\text{ER}_{25}$} & \textbf{$\text{ER}_{40}$} & \textbf{$\text{ER}_{50}$} \\
        \midrule
        CoSPED (ROME) & 1.6 & 0.5 & 0.1 & 0 & 0 \\
        \midrule
        Ethicist (ROME) & 1.5 & 0.4 & 0.1 & 0 & 0 \\
        \midrule
        CoSPED & 85.2 & 78.2 & 70.5 & 66.6 & 63.9 \\
        \midrule
        Ethicist & 84.2 & 78.1 & 70.4 & 65.9 & 61.2 \\
        \bottomrule
    \end{tabular}
    \caption{Privacy Leakage Methods Exact Extraction Rates}
    \label{tab:privacy}
\end{table}

Our ROME method reduces extraction rates from up to 64\% at $\text{ER}_{50}$ to near zero across both attacks. Defense effectiveness increases with longer token-required matches, successfully disrupting autoregressive generation. The minimal changes in $\delta$ Accuracy and LB evaluations confirm that our modified ROME defense selectively targets memorized content without impacting general capabilities.

\section{Conclusion}

In this work, we proposed novel loss functions and decoding strategies that improve extraction performance and enhance the understanding of data leakage risks. Through ablation studies, we identified optimal designs that maximize exact extraction rates, with CoSPED consistently achieving strong results across different metrics, proving the consistency between generated outputs and memorized training data.

Our various analyses highlighted CoSPED's superior performance compared to Ethicist and CLM on GPT-Neo model. Additional experiments on Pythia further validated the universality of CoSPED and indicated that prefix length has a greater impact on extraction performance than soft prompt length. These findings proved the growing risks of targeted data extraction from large language models.

To address this, we applied a modified Rank-One Model Editing method to defend against such extractions. Our defense proved effective against both CoSPED and Ethicist, while preserving the model's general linguistic behavior.

In summary, our work exposed the vulnerability of LLMs to soft prompt extraction attacks and introduced effective defense strategies. We hope this study encourages greater attention to prompt-based privacy risks and inspires future research in robust, model-aware protection techniques.

\section{Limitations and Future Works}

Although we performed multiple model tests, soft prompt requires retraining. One potential improvement is to explore transferability of soft prompt embeddings, referenced in~\cite{kim2023propile, mausblack}.
Specifically, projecting soft embeddings to vocabulary tokens via Euclidean distance~\cite{mausblack} provides theoretical support for universal usage through hard token conversion.

Moreover, developing more robust defense mechanisms to protect LLMs from soft prompt data extraction attacks is also critical for future work. By analyzing how these attacks exploit vulnerabilities in the model's architecture and training data, we hope to identify strategies to mitigate or prevent unauthorized data access, making LLMs more resilient to such attacks in real-world scenarios.

\bibliography{aaai2026}

\appendix

\section{Training Settings}

In this section, we provide details of our training settings and parameters for reference. Our hardware environment is Intel(R) Core(TM) i9-11900K @ 3.50GHz CPU, 3 NVIDIA Corporation GA102 RTX 3090 GPUs, 130GiB of memory, and Ubuntu 22.04.4 LTS.

Soft prompts are utilized following \cite{lester-etal-2021-power} method with randomized initial setup. The testing models are GPT-Neo 1.3B \cite{gpt-neo} and Pythia 1.4B \cite{biderman2023pythia}. For training parameters, we set batch size at 12, initial learning rate at $1 \times 10^{-3}$ with a Cosine Annealing learning rate decay scheduler and a warm-up for 500 steps, AdamW optimizer, and 30 training epochs. Also, the main results utilize soft prompt length at 100 and prefix and suffix length at 50 for comparison.

Smooth loss sets top $N$ at 5. Focal loss contains $\alpha$ at 0.6 and $\gamma$ at 2.0. Dynamic loss has $N$ at 5, $T_D$ at 0.1, and $\alpha$ at 5. Additive loss has $\alpha$ at 0.7, and Common loss has $\beta$ at 0.2. Self Consistency decoding strategy sets $N_\text{SCD}$ at 3, and $\mathcal{D}_{\text{Optimal}}$ at 0.7. The seed is set at 42, and we utilize fp16 for faster training.

In decoding strategy experiments, Top-$P$ \cite{Holtzman2020The} sets $p = 0.7$ and temperature at 0.8. Beam Search sets beam size to 10, and Top-$k$ \cite{fan2018hierarchical} has $k$ at 10. Diverse Beam Search sets group at 3, and beam size at 9.

\section{Test for Transferability}

We tried 2 different methods to realize the soft prompt transferability. One is about \textbf{Transfer} from \cite{mausblack}, which was utilized by \cite{kim2023propile} on their soft prompt transferability usage. The source soft prompt was trained to induce leakage of PII. The other method is about \textbf{Scaling}, which extracts the soft prompt embeddings and scales them to fit the other model's input embedding dimension.

Table \ref{tab:trans} provides our soft prompt transferability testing results. All experiments focus on the GPT-Neo-2.7B model, and the pre-trained soft prompt is from the GPT-Neo-1.3B model, which shares the same decoding vocabulary but a different input embedding size, equaling the soft prompt tokens a similar characteristic.

The transfer method uses Euclidean Distance to calculate the closest hard tokens from the input embeddings, which would be the model's input vocabulary, and uses the closest hard tokens to represent soft prompt embeddings to achieve similar performance. However, from our experiments, the closest hard tokens follow the same as vocabulary, which starts from token "0" and ends at token "99" for a 100-token length soft prompt embedding, which did not successfully show the diversity of soft prompts. Also, the transferred hard tokens did not represent any performance related to extracting more training data, which indicates the soft prompt is not replaceable by the closest hard tokens in the data extraction task. This method was claimed to be successful according to \cite{kim2023propile}, but our task is to generate continuous suffix tokens, which share different features from \cite{kim2023propile}'s short noncontinuous PII information.

According to the theory, the soft prompt scaling method is reasonable because we can scale soft prompt embeddings based on the model input embedding size. Firstly, we calculate the difference in the input embedding size between our trained and larger models. Then, we utilize a projection layer, which is a linear layer, to directly project the embedding tensors from the source dimension to the target dimension. However, the extracted results show that the trained soft prompts lose their distinct characteristics after scaling. This can not induce the model to leak more training data, even providing worse extraction results. One potential reason to explain this is that during the projections, a lot of noise was added in, which could alter the prompt's effect.

\begin{table}[t]
    \centering
    \small
    \begin{tabular}{ccc}
        \toprule
        \multirow{1}{*}{\textbf{Method}} & \multirow{1}{*}{\textbf{$\text{ER}_{50}$}} & \multirow{1}{*}{\textbf{$\text{ER}_{30}$}}\\
        \midrule
        Original & 59.0 & 66.2\\
        Transferred Soft Prompt & 38.4 & 45.1\\
        Scaled Soft Prompt & 35.4 & 41.5\\
        \bottomrule
    \end{tabular}
    \caption{Experiment Results of Soft Prompt Transferability on GPT-Neo-2.7B model}
    \label{tab:trans}
\end{table}

\end{document}